# Network QoS Management in Cyber-Physical Systems

Feng Xia[1,3], Longhua Ma[2], Jinxiang Dong[1], and Youxian Sun[2]
[1]College of Computer Science and Technology, Zhejiang University, Hangzhou 310027, China
f.xia@ieee.org
[2]State Key Lab of Industrial Control Technology, Zhejiang University, Hangzhou 310027, China
lhma@iipc.zju.edu.cn
[3]Faculty of Information Technology, Queensland University of Technology, Brisbane, Australia

**Abstract**

*Technical advances in ubiquitous sensing, embedded computing, and wireless communication are leading to a new generation of engineered systems called cyber-physical systems (CPS). CPS promises to transform the way we interact with the physical world just as the Internet transformed how we interact with one another. Before this vision becomes a reality, however, a large number of challenges have to be addressed. Network quality of service (QoS) management in this new realm is among those issues that deserve extensive research efforts. It is envisioned that wireless sensor/actuator networks (WSANs) will play an essential role in CPS. This paper examines the main characteristics of WSANs and the requirements of QoS provisioning in the context of cyber-physical computing. Several research topics and challenges are identified. As a sample solution, a feedback scheduling framework is proposed to tackle some of the identified challenges. A simple example is also presented that illustrates the effectiveness of the proposed solution.*

## 1. Introduction

In the last two years, a revolutionary transformation from stand-alone, self-contained embedded systems to cyber-physical systems (CPS) [1-3] has commenced. Technical evolutions in sensing, computing, and networking, particularly deeply embedded sensors and wireless sensor/actuator networks (WSANs), are responsible for this tendency. Cyber-physical systems are integrations of computation, networking, and physical dynamics, in which embedded devices are networked to sense, monitor and control the physical world. CPS is an area yet to be explored, with almost all related papers being position papers that discuss the grand challenges and possibilities.

The integration of computing and physical processes is not new. Cyber-physical systems exist today, but in a much smaller scale in size and complexity than the anticipated CPS of the future. The revolution will mainly come from massive networking of embedded computing devices such as sensors and actuators [1]. This revolution will be featured by the envisioned transform that CPS will make on how we interact with the physical world, just like the Internet transformed how we interact with one another. To facilitate unprecedented interactions between human beings and the physical world, networking will become a crucial ingredient due to the need for coupling geographically distributed computing devices and physical elements. It has also become evident that WSANs will be adopted in various CPS to serve as the underlying network infrastructure [4].

In this paper, we are concerned with network design in CPS, with emphasis on network quality of service (QoS) management. A vision of CPS is given in Section 2. The requirements of supporting QoS in WSANs that serve CPS are examined in Section 3. Several research topics of interest and relevant challenges are identified in Section 4. As a sample framework for possible solutions, we propose to exploit the feedback scheduling technology in managing the network QoS in Section 5. An illustrative example is presented with promising results. Finally, concluding remarks are given in Section 6.

## 2. The vision

In a future CPS, as shown in Figure 1, a large number of embedded, possibly mobile computing devices will be interconnected through WSANs, constituting various autonomous subsystems that provide certain services for end users (i.e. human beings). A CPS may be composed of numerous subsystems. Global information sharing is achieved by connecting WSANs to the Internet. For instance, in a cyber-physical city, there may be diverse cyber-physical subsystems for, among others, personal health care, smart home, intelligent transportation, facilities maintenance, and public security. CPS will become pervasive in virtually all fields of science and engineering, such as industry, agriculture, health care, building, military, security, environmental science, biology, and geology, as well as our everyday life.

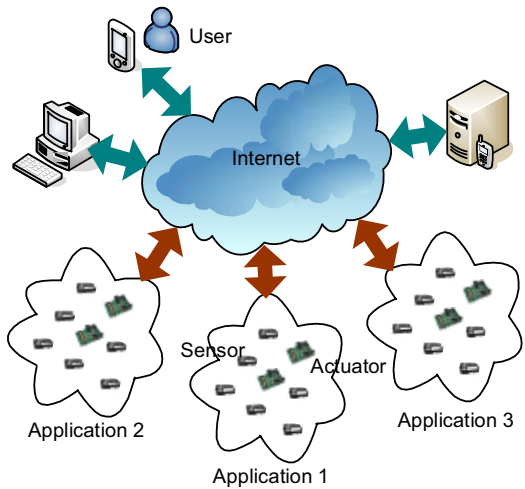

**Figure 1. Physical topology of a CPS**

From an abstract view, WSANs serve CPS as the *interface* between the cyber system and the physical system. Sensors gather information about the physical world, while actuators react to this information by performing appropriate actions upon the physical world. WSANs enable cyber systems to monitor and manipulate the behavior of the physical world. Consequently, the performance and even usability of CPS will heavily depend on the design of these WSANs. With pervasive CPS, we will be able to improve the quality of life through better interacting with the environment in which we live.

## 3. Network characteristics and QoS requirements

Since WSAN design plays a critically important role in building CPS, as mentioned above, we will focus on this issue thereafter. In this section, we will first examine the main features of WSANs and then discuss the requirements of QoS support in this regard.

WSANs are a new generation of sensor networks that feature coexistence of sensors and actuators. While wireless sensor networks (WSNs) are a *passive* information gathering infrastructure, WSANs are *active* in that it enables cyber systems to not only monitor but also manipulate the behavior of physical world [5-8]. Below we discuss some features of WSANs that are closely related to QoS provisioning.

- *Resource constraints*. Sensor nodes are usually low-cost, low-power, small devices that offer only limited data processing capability, transmission rate, battery energy, and memory. In particular, energy conservation is critically important for extending the lifetime of the network. While actuator nodes typically have stronger computation and communication capabilities and more energy budget relative to sensors, resource constraints apply to both sensors and actuators.
- *Platform heterogeneity*. Possibly designed using different technologies and with different goals, sensors and actuators are different from each other in many aspects such as capabilities, functionality, and number. In a large-scale CPS, the hardware and networking technologies used in the underlying WSANs may differ from one subsystem to another.
- *Dynamic network topology*. Node mobility is an intrinsic nature of many applications such as intelligent transportation, assisted living, urban warfare, planetary exploration, and animal control, among others. During runtime, new sensor/actuator nodes may be added or removed; some nodes may even die due to exhausted battery energy. All of these factors cause the network topologies of WSANs to change dynamically.
- *Mixed traffic*. Diverse applications may need to share the same WSAN that induce both periodic and aperiodic data. This feature will become increasingly common as the scale of WSANs grows. Furthermore, sensors for different kinds of physical variables, e.g., temperature, humidity, location, and speed, generate traffic flows with different characteristics (e.g. message size and sampling rate).

CPS by nature is application-oriented. Therefore, WSANs have to provide QoS support so as to satisfy the service requirements of target applications. From an end user's perspective, real-world WSAN applications have their specific requirements on the QoS of the underlying network infrastructure [4,8]. For instance, in a fire handling system, sensors need to report the occurrence of a fire to actuators in a timely and reliable fashion; then, the actuators equipped with water sprinklers will react by a certain deadline so that the situation will not become uncontrollable.

Different applications may have different QoS requirements. For instance, for a safety-critical control system, large delay in transmitting data from sensors to actuators and packet loss occurring during the course of transmission may not be allowed, while they may be acceptable for an air-conditioning system that maintains the temperature insider an office.

Although QoS is an overused term, there is no common definition of this term. Conceptually, it can be regarded as the capability to provide assurance that service requirements of applications can be satisfied. Depending on the application, QoS in WSANs can be characterized by reliability, timeliness, robustness, availability, and security, among others. Some QoS parameters may be used to measure the degree of satisfaction of these services, such as throughput, delay, jitter, and packet loss rate.

## 4. Possibilities and challenges

Over the years, in order to meet the requirements of diverse applications on network QoS, significant attempts have been made to provide end-to-end QoS support at different network protocol layers. Particularly, Internet QoS has been a focus of enormous research and development activities [9]. Due to the above-described characteristics of WSANs, however, existing QoS mechanisms may not applicable to WSANs [10,11], particularly in the context of cyber-physical computing. Supporting QoS in WSANs is an area nearly unexplored. In this section, we identify some research topics of interest as well as related challenges that need to be overcome.

### 4.1. Service-oriented architecture

The concept of service-oriented architecture (SOA) [12] is not new and has been widely used in, for example, the web services domain. However, many of its elegant potentials have not ever been explored in CPS, though SOA will undoubtedly have a major impact in many branches of technology [6,13]. A general interpretation of SOA is that it is an architectural style encompassing a set of services for building complex systems of systems. It can be regarded as a model in which functionality is decomposed into small, distinct units, i.e. services. As an architectural evolution and a paradigm shift in systems integration, SOA enables rapid, cost-effective composition of interoperable, scalable systems based on reusable services exposed by these systems. This is particularly useful for QoS provision in WSANs that are integrated into large-scale, highly complex CPS.

Identifying and specifying services are crucial for exploiting SOA in WSAN design. A large number of questions need to be answered in this respect. For example, how many categories of services should be classified? What are the functionality, interface, and properties of each service? What are its quality levels relevant to performance requirements, from the perspective of network QoS management? In particular, how to deal with the difference between sensors and actuators when specifying services?

### 4.2. QoS-aware communication protocols

Since network performance depends, to a large extent, on the underlying communication protocols used, communication protocols are one of the most widely-studied aspects of sensor networks. While the protocol stack generally consists of the physical layer, the data link layer, the network layer, the transport layer, and the application layer, in the literature special attention has been given to medium access control (MAC), routing, and transport protocols. There are also some attempts to develop communication protocols that provide QoS support in WSNs, but QoS-aware MAC, routing, and transport protocols for WSANs are still open issues [14].

In order to efficiently support QoS in WSANs, communication protocols need to be designed with in mind the platform heterogeneity, specifically the heterogeneity between sensors and actuators that are involved in the CPS. For this reason, state-of-the-art QoS-aware MAC, routing, and transport protocols developed for WSNs may not be suitable for WSANs.

As an essential component of QoS, service differentiation should be supported by communication protocols. A CPS may encompass diverse applications, which may differ significantly in their QoS requirements. The best-effort service offered by current networking technologies cannot provide different QoS to different applications. Therefore, communication protocols for WSANs should be designed to perceive the service requirement of each type of traffic so that it can be guaranteed a specific service level. In practice, however, the best-effort service is likely to be the standard for the foreseeable future [9]. It is therefore necessary for new QoS mechanisms to be layered on top of the existing networks.

Cross-layer design has proved to be effective in optimizing the network performance and hence may be incorporated in the development of QoS-aware network communication protocols for CPS. Much work can be conducted in this line. For example, the prioritization of traffic at lower layers might be associated with the application performance at the application layer.

### 4.3. Resource self-management

Resource management is of paramount importance for QoS provision because the resource budgets need to be guaranteed in order to achieve certain QoS levels. This is particularly true for WSANs where computing, communication and energy resources are inherently limited [15]. Generally speaking, a higher level of QoS corresponds to a need of more resources, e.g. CPU time, memory size, bandwidth and/or energy. Resource management in WSANs is challenging, because of the ever-increasing complexity of CPS, highly dynamic feature of the networks, and changing and unpredictable environments.

To overcome these challenges, self-management technologies [16] are needed. This implies that the system needs to address resource management issues in an autonomous manner. With respect to changes in resource availability, resource manager will automatically adapt resource usage in a way that the resulting overall QoS is optimized. This has to be conducted in an efficient way. Since the resources are limited, the overhead of resource management should be minimized. To maintain scalability, furthermore, distributed mechanisms have to be explored. In Section 5, we will propose a promising framework to achieve resource self-management.

### 4.4. QoS-aware power management

Energy conservation is a major concern in WSANs [17]. The lifetime of untethered sensor/actuator nodes is tightly restricted by the available battery energy. Since wireless communication is much more energy-expensive than sensing and computation, the transmission power of nodes has to be properly managed in a way that the energy consumption is minimized. However, minimizing energy consumption and maximizing QoS are in most cases two conflicting requirements. For instance, reliability can be improved by increasing the number of allowed retransmissions or using higher transmission power levels; however, more energy will be expended in both cases. Therefore, tradeoffs must be made between energy conservation and QoS optimization. The problem then becomes how to make these tradeoffs at runtime. Is it possible to find an integrated performance metric that covers both energy efficiency and QoS, and then optimize it?

Depending on the network topology and the QoS requirements, the transmission power management mechanisms for actuator nodes may be different from those used in sensor nodes. Thus the QoS can be maximized through exploiting the different capabilities of sensors and actuators. In like manner, different transmission power levels may be assigned to the same node with respect to different QoS requirements imposed by different types of traffic. In-network computation can be exploited to reduce the energy consumption of both sensor and actuator nodes since it reduces traffic load at the cost of slightly increased computation in each involved node. Still, the inherently non-deterministic and open nature of wireless channels poses large challenges for power management subject to QoS constraints.

## 5. Feedback scheduling as a solution

The challenges identified above are formidable. This section will propose a feedback scheduling framework as a sample solution for overcoming some of the challenges identified in Section 4.3.

Taking advantage of the well-established feedback control theory and technology, feedback scheduling offers a promising approach to autonomous resource management in dynamic and unpredictable environments. Previous work has showed that feedback scheduling is capable of handling uncertainties in resource availability by automatically adapting to dynamic changes [18-21].

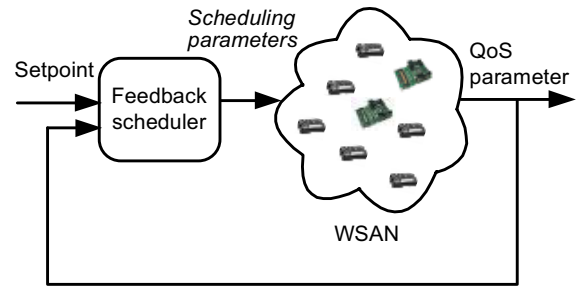

**Figure 2. Feedback scheduling framework**

A general feedback scheduling framework that can be used to achieve QoS support in WSANs through resource self-management is given in Figure 2. The basic role of the feedback scheduler is dynamically adjusting specific scheduling parameters of relevant traffic so as to maintain a desired QoS level. In control terms, the system output, i.e. a certain QoS parameter, is the *controlled variable*, the adjustable scheduling parameters are the *manipulated variable*, and the desired value of the QoS parameter is the *setpoint*. The feedback scheduler can normally be designed using control theory and technology. In this way, the predictability of the CPS can be enhanced. In the presence of unpredictable changes in resource availability, for example, an increase in traffic load, a well-designed feedback scheduler will always drive the resulting QoS to the pre-determined level. This mechanism can be implemented separately in each node and hence is scalable.

## 5.1. An illustrative example

To illustrate the feasibility of the above scheme, consider a simple WSAN [8] as shown in Figure 3, where $s_1$, $s_2$, $s_3$, and $s_4$ are source (sensor) nodes, $s_5$ is an interfering source node, $s_6$ is an intermediate node, $a_1$ and $a_2$ are actuator nodes. These nodes have to compete for the use of the same wireless channel for data transmission. The utilized communication protocol is ZigBee with a data rate of 250 kbps. All data packets transmitted over the network are 45 bytes in size, which may correspond to a payload of 32 bytes and an overhead of 13 bytes. The default sampling period for each source node is 10 ms. The sampling period of $s_5$ cannot be adjusted at runtime because it is an interfering node. The deadline of a data packet is assumed to be equal to current sampling period of the relevant source node.

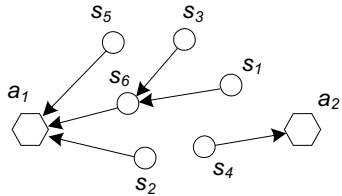

**Figure 3. A simple WSAN**

The system runs as follows. At the beginning, all nodes except $s_3$, $s_4$, and $s_5$ are active; $s_5$ is switched on at time t = 20s and off at time t = 40s; $s_3$ and $s_4$ remains off until t = 60s. In real-time CPS, the deadline miss ratio (DMR) associated with the data transmission from each source node ($s_1$, $s_2$, $s_3$, and $s_4$) to the actuator is of the major concern.

Figure 4 depicts the deadline miss ratio for each sensor when there is no QoS management mechanism. It is clear that the deadline miss ratios change dramatically as the traffic load over the network varies. The average deadline miss ratio throughout the simulation is as high as 81.1%, 58.4%, 100%, and 98.5%, respectively, for each source node.

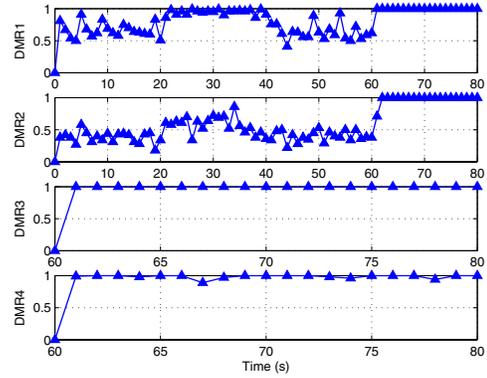

**Figure 4. Deadline miss ratios without QoS management**

To address this problem, a separate feedback scheduler for each source sensor is designed using fuzzy logic control technology. The interested reader is referred to [8] for details on the design of this feedback scheduler. It is omitted here due to space limitation. The controlled variable is the deadline miss ratio and the manipulated variable is the sampling period of the sensor. The feedback scheduler executes in a periodic manner, with an invocation interval of 1s. The desired deadline miss ratios for all sensors are set to 10%.

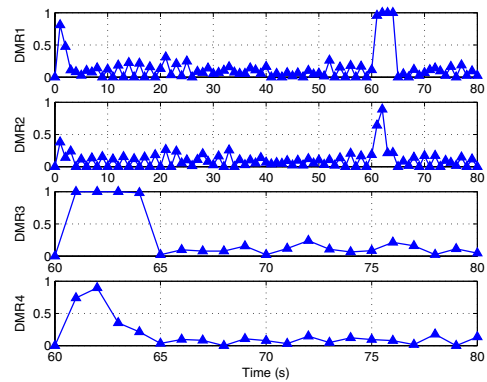

**Figure 5. Deadline miss ratios with QoS management**

Figure 5 shows the deadline miss ratios when the designed feedback scheduler is used. It can be seen that the deadline miss ratios for all data transmissions keep around the desired level 10% and are much lower than those without QoS management (Figure 4) almost all the time. The average deadline miss ratios for four source sensors are 14.2%, 10.5%, 24.2%, and 14.2%, respectively, which are significantly lower than those in the previous case.

From these results, it is summarized that the feedback scheduling technology can be used in WSANs to realize resource self-management and to provide network QoS guarantees in CPS. However, much work is left to be done in order to make it practically useful.

## 6. Concluding remarks

CPS is opening up unprecedented opportunities for research and development in numerous disciplines. Meanwhile, researchers currently face a large number of challenges that need to be overcome before the envisioned CPS become reality. This paper has examined the network QoS management issue in CPS. In order to spark new interests in this line, several research topics and challenges in supporting QoS in WSANs have been identified. A feedback scheduling framework has also been proposed as a sample solution to network QoS management in CPS. It is argued that WSAN QoS management will play an important role in CPS design and implementation. Extensive research is expected in this exciting new area.

## Acknowledgments

The first author would like to thank A/Prof Yu-Chu Tian at QUT, Australia, for helpful discussions. This work is supported in part by China Postdoctoral Science Foundation under grant No. 20070420232, Natural Science Foundation of China under grant No. 60474064, Zhejiang Provincial Natural Science Foundation of China under grant No. Y107476, and Australian Research Council (ARC) under Discovery Projects grant No. DP0559111.